\documentclass[twocolumn,aps,prl,amsmath,floatfix]{revtex4}
\usepackage{graphicx}
\begin{document}

\title{Subband filling and Mott transition in Ca$_{2-x}$Sr$_x$RuO$_4$}
\author{A. Liebsch$^1$ and H. Ishida$^2$}
\affiliation{
\mbox{
$^1$Institut f\"ur Festk\"orperforschung,~Forschungszentrum J\"ulich,
        ~52425 J\"ulich, Germany} \\
\mbox{
$^2$College of Humanities and Sciences, Nihon University, and CREST JST,
        ~Tokyo 156, Japan}\\
}

\begin{abstract}
A new concept is proposed for the paramagnetic metal insulator transition 
in the layer perovskite Ca$_{2-x}$Sr$_x$RuO$_4$. Whereas the pure Sr 
compound is metallic up to very large Coulomb energies due to strong 
orbital fluctuations, structural changes induced by doping with Ca give 
rise to a interorbital charge transfer which makes the 
material extremely sensitive to local correlations. Using dynamical mean 
field theory based on finite temperature multi-band exact diagonalization 
it is shown that the combination of crystal field splitting and onsite 
Coulomb interactions leads to complete filling of the $d_{xy}$ band and 
to a Mott transition in the half-filled $d_{xz,yz}$ bands.\\ \\ 
DOI: \hfill PACS numbers: 71.20.Be, 71.18.+y.
\end{abstract}
\maketitle

The layer perovskite Ca$_{2-x}$Sr$_x$RuO$_4$ has attracted wide interest
during recent years because of the complex sequence of electronic and 
magnetic phases which arise when Sr is iso-electronically substituted by Ca
\cite{nakatsuji,wang,lee}. 
While the pure Sr compound exhibits unconventional superconductivity
\cite{maeno}, 
the structural distortions induced by the smaller Ca ions ultimately 
lead to an antiferromagnetic Mott insulator. The physical mechanism of how 
this multi-band material with four electrons per Ru ion evolves from the 
metal Sr$_2$RuO$_4$ towards the insulator Ca$_2$RuO$_4$ is presently 
not well understood.   

According to band structure calculations \cite{oguchi} the $t_{2g}$ orbitals
in Sr$_2$RuO$_4$ are approximately equally occupied. Because of the planar
geometry, these bands split into a wide, nearly two-dimensional $d_{xy}$ band
and two narrow, nearly one-dimensional $d_{xz,yz}$ bands. Clearly, the Mott 
transition in such a highly anisotropic system cannot be understood in terms
of single-band models, or multi-band extensions assuming degenerate subbands. 
To analyze the role of local Coulomb interactions in Ca$_{2-x}$Sr$_x$RuO$_4$ 
several aspects were investigated within dynamical mean field theory (DMFT) 
\cite{dmft}, using a variety of impurity solvers to treat onsite correlations 
\cite{liebsch,anisimov,EPL,sakai,pchelkina,dai}. 

While it is generally agreed upon that Coulomb interactions in Sr$_2$RuO$_4$ 
lead to a sizeable band narrowing and effective mass enhancement, and a shift 
of the $d_{xy}$ van Hove singularity towards the Fermi level due to $d_{xz,yz} 
\rightarrow d_{xy}$ charge transfer \cite{liebsch,EPL,sakai,pchelkina}, the 
Mott transition near the Ca end of the phase diagram is complicated 
because of the charge rearrangement among $t_{2g}$ orbitals when the Ca/Sr 
concentration is varied. Using the non-crossing approximation \cite{nca}, Anisimov 
{\it et al.}~\cite{anisimov} obtained successive, `orbital selective' Mott 
transitions upon increasing the onsite Coulomb energy $U$: first for the narrow 
$d_{xz,yz}$ bands and subsequently for the wide $d_{xy}$ band. These transitions 
arise via a $d_{xy}\rightarrow d_{xz,yz}$ charge transfer, i.e., 
$(n_{xy},n_{xz},n_{yz})\approx(2/3,2/3,2/3)\rightarrow(0.5,0.75,0.75)$. 
Orbital selective Mott
transitions, with the same interorbital charge transfer, were found  
also by Dai {\it et al.}~\cite{dai} within slave boson mean field calculations 
for a three-band model consisting of wide and narrow semi-circular densities of 
states. To account for the Ca induced octahedral distortions \cite{fang}, the 
$d_{xy}$ band was assumed to be narrower than the $d_{xz,yz}$ bands. 
Accordingly, in contrast to Ref.~\cite{anisimov}, $d_{xy}$ is the first band 
to become insulating with increasing $U$. An important parameter in this work 
is the crystal field splitting $\Delta$ between $t_{2g}$ orbitals, which yields 
Mott transitions at much smaller values of $U$ than for $\Delta=0$. In both 
models, a second interorbital charge transfer due to additional structural 
modifications, i.e., 
$(n_{xy},n_{xz},n_{yz})=(0.5,0.75,0.75)\rightarrow(1.0,0.5,0.5)$, is required 
in the limit $x\rightarrow0$
to yield the antiferromagnetic insulating properties of Ca$_2$RuO$_4$
\cite{anisimov}.

In the present work, we use finite temperature exact diagonalization (ED) DMFT 
\cite{ed} to study the nature of the Mott transition in Ca$_{2-x}$Sr$_x$RuO$_4$. 
Since a $t_{2g}$ tight-binding Hamiltonian including the full complexity of the 
octahedral distortions as a function of Ca/Sr concentration is not yet available, 
we use the Sr$_2$RuO$_4$ density of states components as a single-particle starting 
point (see Fig.~1). As shown by Fang {\it et al.}~\cite{fang}, a key effect caused 
by rotation, tilting and flattening of oxygen octahedra is the {\it increasing} 
$d_{xy}$ orbital occupancy with increasing Ca concentration: In the limit $x=0$
(pure Ca), the spin-averaged occupancy is about $n_{xy}\approx 0.83$,  
compared to $n_{xy}\approx 0.64$ for $x=2$ (pure Sr). To account for this 
charge transfer, we allow, as in Ref.~\cite{dai}, for a crystal field 
splitting $\Delta$ between $d_{xy}$ and $d_{xz,yz}$ states. For instance, a 
lowering of the $d_{xy}$ bands by $\Delta=0.2\ (0.4)$~eV yields 
$n_{xy}\approx 0.74\ (0.83)$. The total occupancy of the $t_{2g}$ bands is four, 
independently of Ca/Sr concentration. 

The main result of this work is a new mechanism for the Mott transition in this 
multi-band system: For realistic values of the Coulomb energy $U$ and crystal 
field splitting $\Delta$, we find a gradual filling of the $d_{xy}$ subband with 
increasing $U$. Once the $d_{xz,yz}\rightarrow d_{xy}$ charge transfer is complete, 
i.e., $(n_{xy},n_{xz},n_{yz})\rightarrow(1.0,0.5,0.5)$, a Mott transition takes 
place in the remaining half-filled $d_{xz,yz}$ bands. The critical Coulomb energy 
at which this transition occurs is lower for larger crystal field $\Delta$.
     
This scenario is differs qualitatively from the ones proposed in 
Refs.~\cite{anisimov,dai}: 
Instead of successive transitions in the half-filled $d_{xy}$ band and 
in the $3/4$-filled $d_{xz,yz}$ bands, we find a common $d_{xy}$ band filling
and $d_{xz,yz}$ Mott transition. The $d_{xy}$ band filling, induced by the 
combined effect of crystal field and Coulomb interactions, signifies a change 
from a correlated metallic state to a correlated band insulating state and does 
not correspond to a Mott transition. On the other hand, since a full $d_{xy}$ 
band implies half-filled $d_{xz,yz}$ bands, the latter readily undergo a standard 
metal insulator transition. Thus, the present three-band system does not exhibit 
orbital selective Mott transitions. Our solution solely relies on the structure
induced increase in $d_{xy}$ occupancy \cite{fang} and is consistent with the
$(n_{xy},n_{xz},n_{yz})=(1.0,0.5,0.5)$ configuration obtained for Ca$_2$RuO$_4$
within LDA+U \cite{anisimov}. Thus, we do not require the more complex path
$n_{xy}\approx 2/3\rightarrow 1/2\rightarrow 1$ as Sr is replaced by Ca.

\begin{figure}[t!]
  \begin{center}
  \includegraphics[width=5.0cm,height=8.0cm,angle=-90]{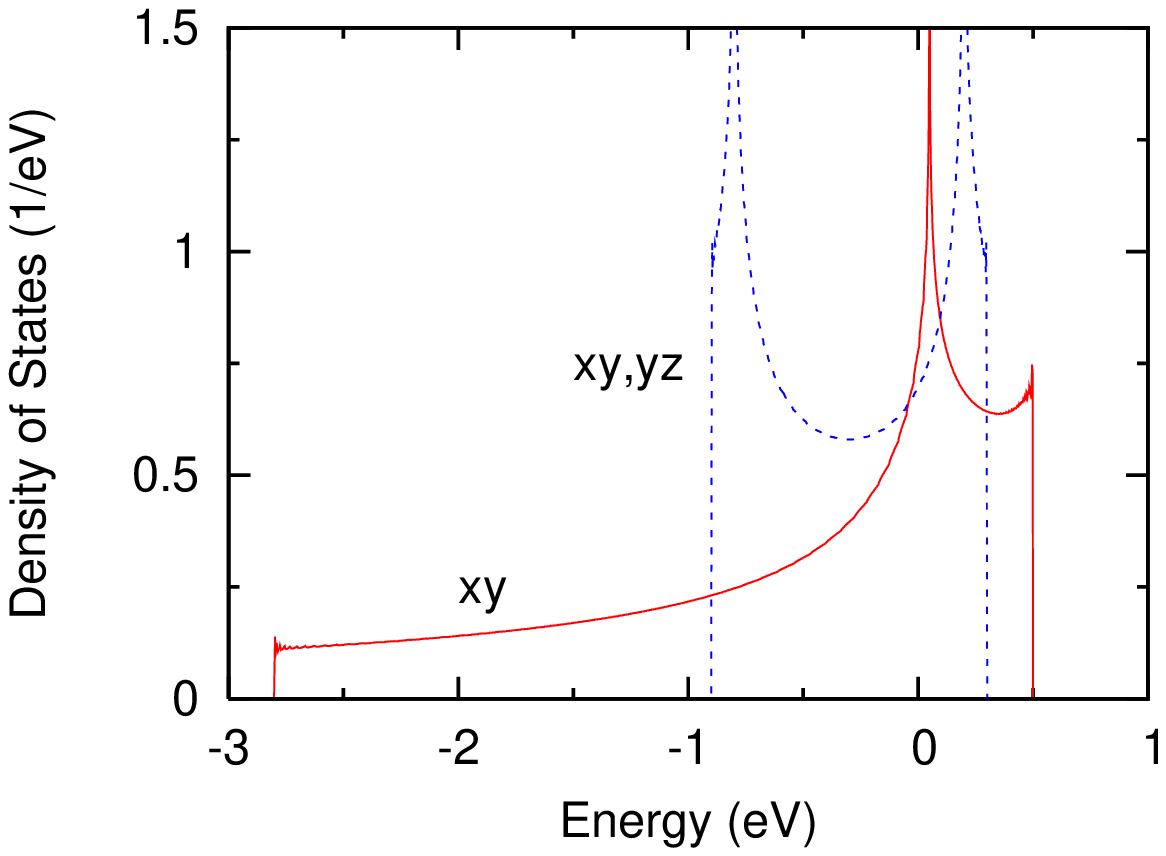}
  \includegraphics[width=5.0cm,height=5.5cm,angle=-90]{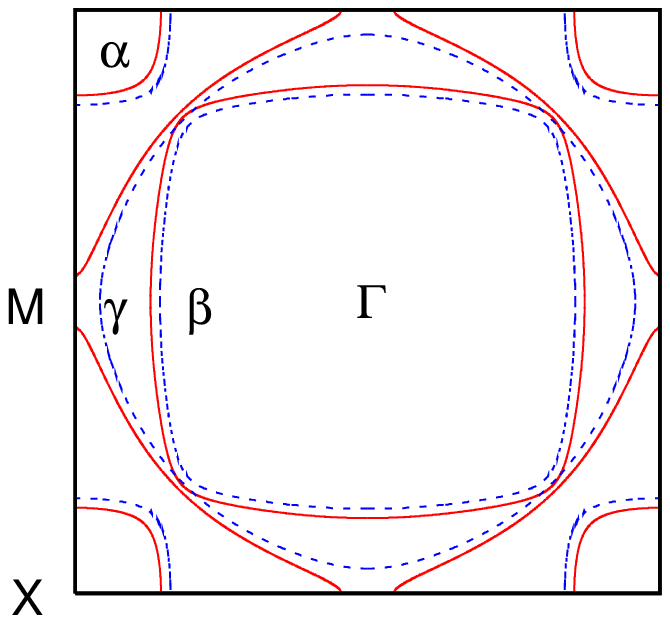}
  \end{center}
  \vskip-2mm
\caption{
Upper panel: Density of states of Sr$_2$RuO$_4$. Solid (red) curve:
wide $d_{xy}$ band; dashed (blue) curve: narrow $d_{xz,yz}$ bands ($E_F=0$).
Lower panel: Fermi surface (schematic) of Sr$_2$RuO$_4$ (dashed blue curve)
and  Ca$_{1.5}$Sr$_{0.5}$RuO$_4$ (solid red curve). 
In the former (latter) case, the $d_{xy}$ van Hove singularity at $\rm M$ 
is above (below) $E_F$.
}\end{figure}

The details of our multi-band ED approach are given in Ref.~\cite{ed}. 
Exploiting the sparseness of the impurity Hamiltonian, much larger clusters 
representing the bath can now be handled, thereby significantly reducing finite 
size effects. All calculations are carried out for full Hund exchange $J$. 
To obtain the orbital occupancy $n_i$ as a function of $U$, for a particular 
$\Delta$, we assume $J=U/4$. The interorbital Coulomb energy is $U'=U-2J$. 
To achieve fast convergence we use the cluster size $n_s=9$ (2 bath levels 
per impurity orbital); for greater precision specific points are calculated 
with $n_s=12$. The temperature is assumed to be $T=20$~meV. We focus here 
on the paramagnetic Mott transition. 

Fig.~1 shows the $t_{2g}$ density of states components for Sr$_2$RuO$_4$,
derived from a tight-binding Hamiltonian \cite{liebsch} fitted to the LDA band 
structure of Ref.~\cite{oguchi}. The band extrema were adjusted slightly in order 
to accomodate four electrons below $E_F$. The $d_{xy}$ van Hove singularity 
lies about 60~meV above $E_F$. The lower panel shows the Fermi surface. 
The $\gamma$ sheet corresponds to $d_{xy}$, the $\alpha$ and $\beta$ sheets
to $d_{xz,yz}$. The latter do not cross because of a small hybridization term. 
Also shown is the Fermi surface for a crystal field $\Delta=0.1$~eV. 
Since the $d_{xy}$ van Hove singularity now lies below $E_F$, the $\gamma$ sheet
has turned from electron-like to hole-like.

\begin{figure}[t!]
  \begin{center}
  \includegraphics[width=7cm,height=8cm,angle=-90]{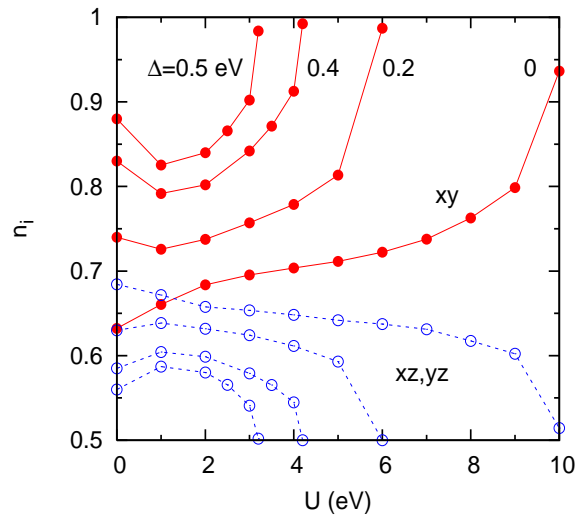}
  \end{center}
  \vskip-2mm
\caption{
Orbital occupancy $n_i$ as a function of Coulomb energy for various crystal 
field splittings $\Delta$, derived within ED/DMFT for $T=20$~meV. 
Solid (red) dots: $n_{xy}$, empty (blue) dots: $n_{xz,yz}$.   
The lines are guides to the eye.
}\end{figure}
 
Fig.~2 summarizes the variation of the $t_{2g}$ orbital occupancies with 
onsite Coulomb energy $U$. In the case of Sr$_2$RuO$_4$ ($\Delta=0$), a 
gradual charge transfer from $d_{xz,yz}$ to $d_{xy}$ states is found, in 
agreement with the trend found within QMC/DMFT for small $U$ \cite{liebsch}. 
The system is seen to remain metallic up to very large $U$.
This result is consistent with previous calculations \cite{EPL} which
revealed no Mott transition up to $U=4$~eV. Near $U_c\approx10$~eV the
interorbital charge transfer is complete with $n_{xy}=1$. The remaining
half-filled $d_{xz,yz}$ bands then undergo a Mott transition (see below). 

Note that the filling of the $d_{xy}$ band for $\Delta=0$ accelerates 
towards increasing $U$. The reason for this trend is that the larger 
$d_{xy}$ occupancy reduces the intra-$t_{2g}$ screening of $U$.
Since at the same time the occupancy of the $d_{xz,yz}$ bands approaches 
one-half, the tendency for the latter bands to undergo a Mott transition
becomes rapidly more favorable. In the absence of the $d_{xy}$ band, the 
half-filled $d_{xz,yz}$ bands (width $W=1.2$~eV) would exhibit a common 
metal insulator transition at $U_c=1.4$~eV, i.e., far below $U=10$~eV 
($U_c\approx0.8\,W$ for twofold degenerate semicircular bands and $J=U/4$ 
\cite{prb70}). Thus the results for $\Delta=0$ may be interpreted as 
$d_{xz,yz}$ Mott transition delayed by strong orbital fluctuations.  

In view of this picture it is plausible that, at finite Ca concentrations, 
the greater initial occupancy of the $d_{xy}$ band associated with $\Delta>0$
gives rise to complete $d_{xy}$ filling and a $d_{xz,yz}$ Mott transition at 
progressively lower values of $U_c$. This is confirmed by the 
results shown in Fig.~2. According to Fang {\it et al.}~\cite{fang} the 
crystal field splitting for Ca$_2$RuO$_4$ induced by octahedral distortions 
is about $0.4$~eV. The $d_{xy}$ band filling and $d_{xz,yz}$ Mott transition 
is then shifted to $U_c\approx4$~eV. Thus, the reduced orbital fluctuations 
greatly diminish the delay of the Mott transition in the half-filled 
$d_{xz,yz}$ bands. A slightly lower $U_c$ would be obtained if other 
effects neglected so far, such as distortion induced band narrowing and 
interorbital hybridization \cite{fang}, are taken into account.       

In the range $0.5\le x \le 2.0$ the octahedral distortions consist mainly
of rotations about the $z$-axis. The associated lowering of the $d_{xy}$ 
band is rather small, giving $\Delta\le 0.2$~eV \cite{fang}. According 
to the results shown in Fig.~2, the correlation induced $d_{xy}$ filling 
and $d_{xz,yz}$ Mott transition then occur at $U_c\approx 6$~eV.
In this doping region the system therefore remains metallic.
This result is consistent with angle resolved photoemission data \cite{wang} 
and optical data \cite{lee} which show that near $x=0.5$ all $t_{2g}$ states 
are itinerant. 

A crystal field $\Delta=0.1\ldots0.2$~eV is large enough to push 
the $d_{xy}$ van Hove singularity below $E_F$. This also agrees 
with the photoemission results \cite{wang} which yield nearly the same 
Fermi surface for Ca$_{1.5}$Sr$_{0.5}$RuO$_4$ and Sr$_2$RuO$_4$, except that,
because of the lowering of the van Hove singularity, the $\gamma$ sheet 
has changed from electron-like to hole-like, as indicated in Fig.~1. These 
data are naturally explained by the results shown in Fig.~2, whereas they 
are difficult to reconcile with the trend towards half-filled $d_{xy}$ bands 
proposed in Refs.~\cite{anisimov,dai}. 
 
 \begin{figure}[t!]
  \begin{center}
  \includegraphics[width=4.5cm,height=8cm,angle=-90]{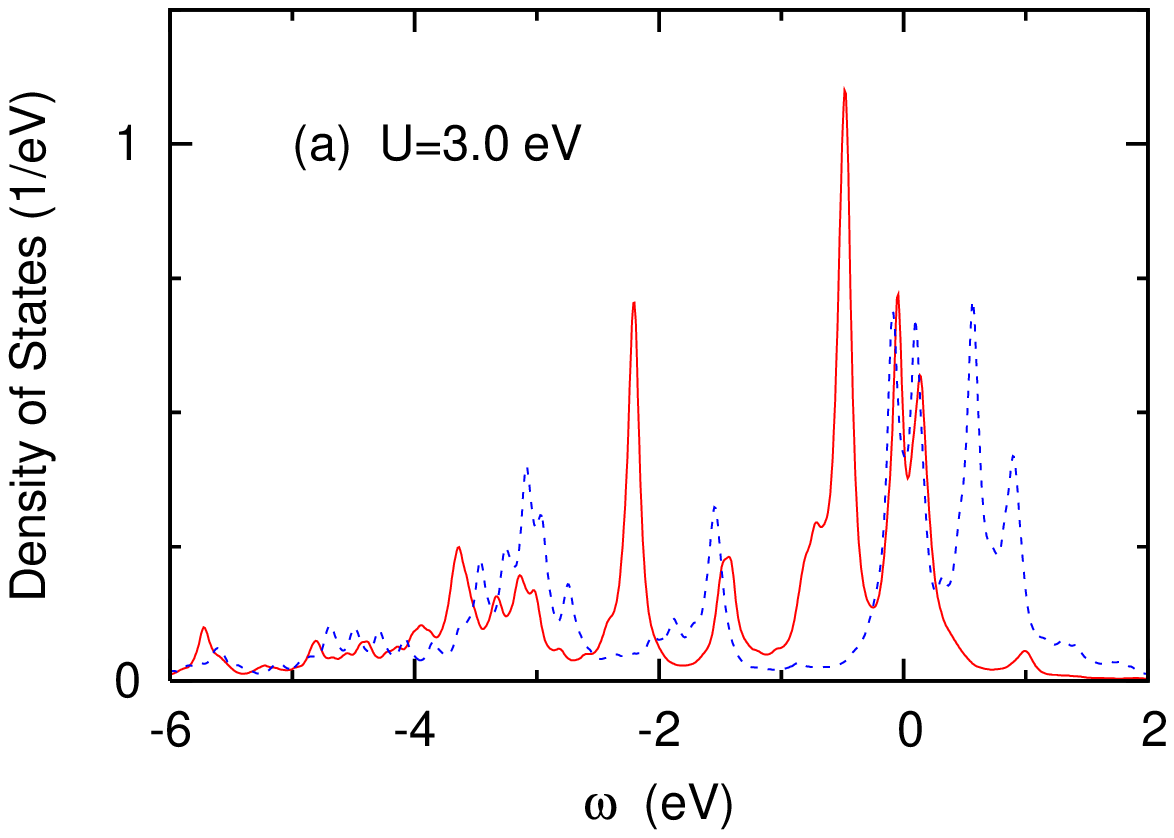}
  \includegraphics[width=4.5cm,height=8cm,angle=-90]{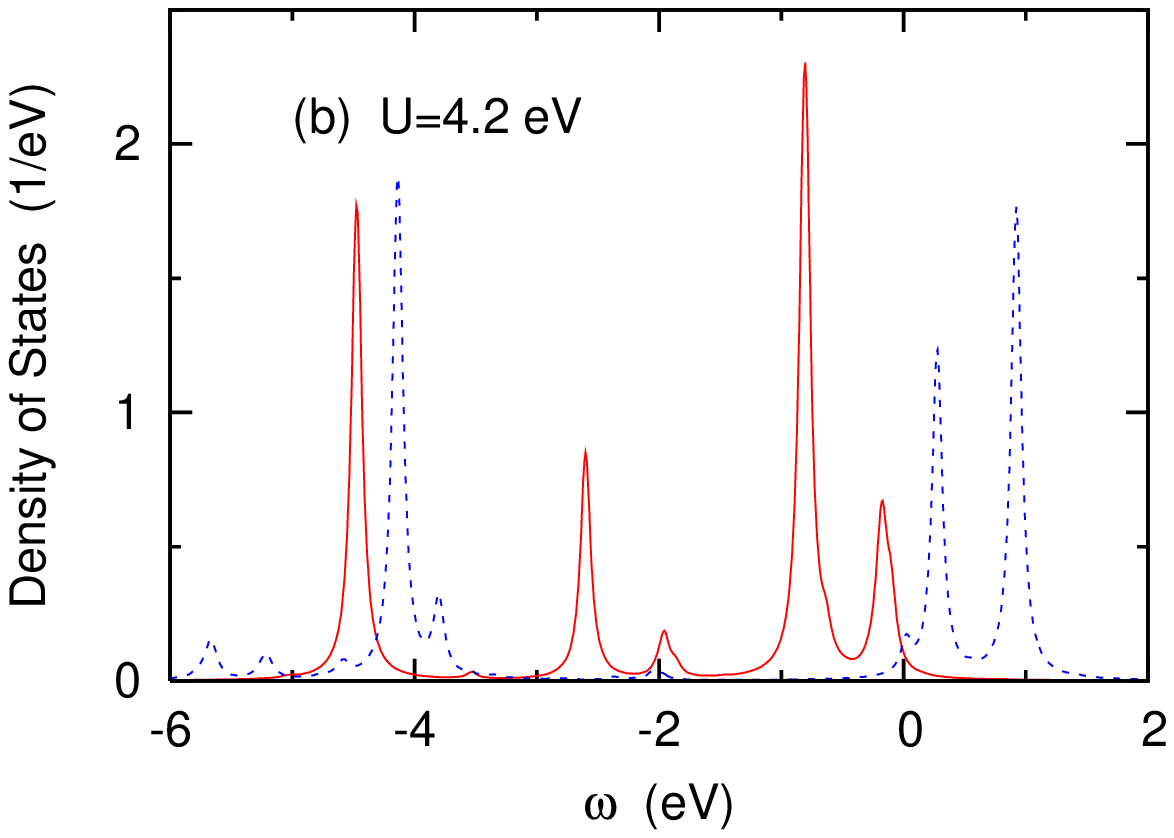}
  \includegraphics[width=4.5cm,height=8cm,angle=-90]{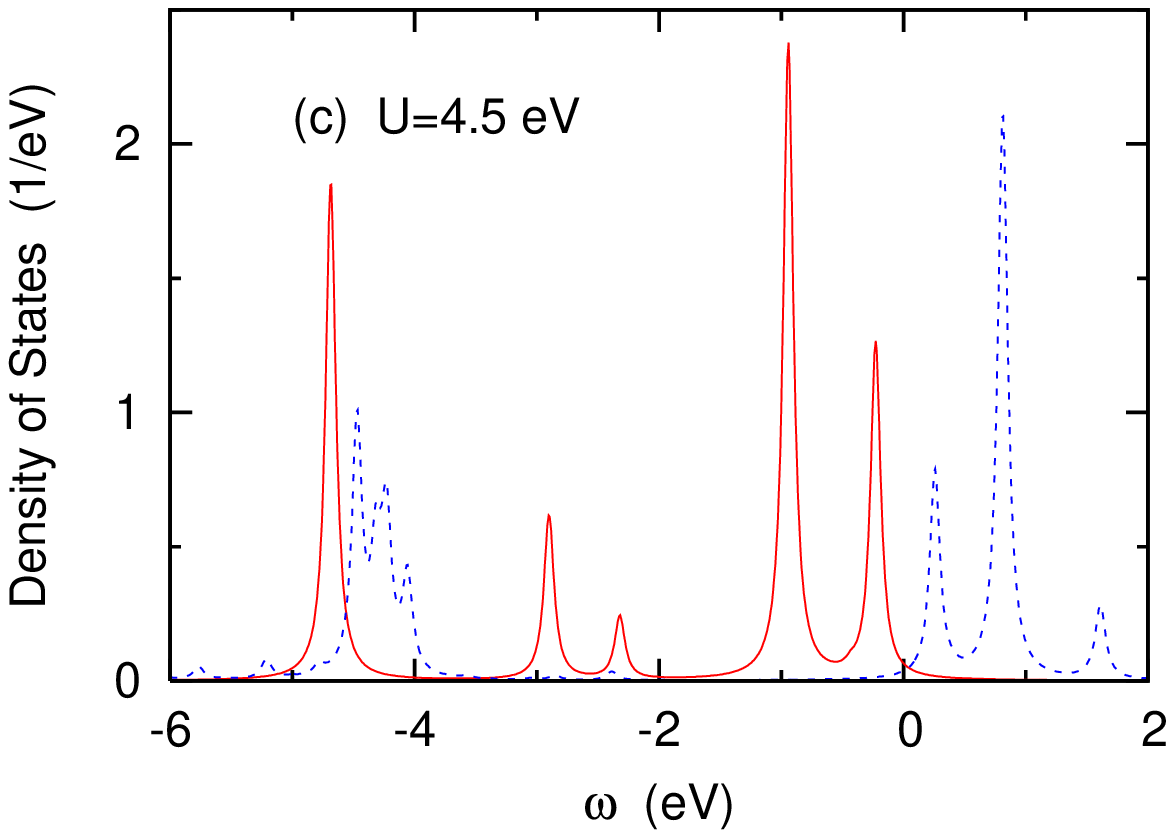}
  \end{center}
  \vskip-2mm
\caption{
Quasi-particle spectra calculated using ED/DMFT for crystal field 
splitting $\Delta=0.4$~eV; $T=20$~meV.
(a) metallic region $U=3.0$~eV; (b) transition near $U_c\approx4.2$~eV;
(c) insulating region $U=4.5$~eV.
Solid (red) curves: $d_{xy}$ band, dashed (blue) curves: $d_{xz,yz}$ bands. 
}\end{figure}

To prove that $d_{xy}$ band filling and $d_{xz,yz}$ Mott transition coincide   
we show in Fig.~3 the $t_{2g}$ quasi-particle spectra for $\Delta=0.4$~eV. To 
avoid uncertainties stemming from the extrapolation from Matsubara frequencies 
to real frequencies we give here the spectra of the cluster Green's functions, 
$A_i(\omega)=-\frac{1}{\pi}{\rm Im}\,G_i(\omega+i\delta)$, with $\delta=50$~meV. 
While spectral details differ from those of the solid \cite{ed}, 
the cluster results are adequate for the distinction 
between metallic and insulating behavior. The spectra for $U=3.0$~eV reveal 
that all subbands are metallic and exhibit appreciable spectral 
weight below the single particle bands, associated with lower Hubbard bands. 
According to Fig.~2, the occupancies are $n_{xy}=0.84$ and $n_{xz,yz}=0.58$.
In contrast, at $U=4.5$~eV the $d_{xy}$ band is filled and the half-filled
$d_{xz,yz}$ bands exhibit a clear separation into upper and lower Hubbard bands. 
The transition between these two regions occurs at $U_c\approx4.2$~eV.
Note that the insulating gap arises between the filled $d_{xy}$ band and the
upper Hubbard bands of the half-filled $d_{xz,yz}$ bands. The quasi-particle
properties near this transition will be discussed elsewhere.

We point out that, in general, the $d_{xy}$ band filling and $d_{xz,yz}$ 
Mott transition do not need to take place at the same $U$. For instance, if
the $d_{xz,yz}$ bands were much wider, their Mott transition would occur
above the $d_{xy}$ band filling. Also, a stronger crystal field could give 
a $d_{xy}$ filling at very small $U$, and a Mott transition in the half-filled 
$d_{xz,yz}$ bands at a larger $U$. Because of the small width of these bands 
the trend seen in Fig.~2 suggests that this possibility should arise only for 
$\Delta>0.5$~eV. An analogous effect was  
discussed by Manini {\it et al.}~\cite{manini} for a two-band model with 
equal bands of semi-circular density of states, off-set via a crystal field 
$\Delta$. For unit total occupancy and small $\Delta$, 
a single transition was found where one band is pushed above $E_F$ and the 
other (then half-filled) exhibits a Mott transition. For larger $\Delta$, 
one band is emptied at small $U$ while the second remains metallic up to a 
Mott transition at larger $U$. 

We also note that the Mott transitions in the $3d^1$ perovskites
LaTiO$_3$ and YTiO$_3$ studied by Pavarini {\it et al.}~\cite{pavarini} 
reveal an almost complete emptying of two $t_{2g}$ subbands, and a 
metal insulator transition in the remaining nearly half-filled band.
Evidently, in all of these multi-band systems, the Mott transition is 
made feasible by a striking suppression of orbital fluctuations.    

In the work discussed above the $d_{xz,yz}$ bands are degenerate. In real
Sr$_2$RuO$_4$ these orbitals interact weakly, giving slightly different 
subband densities of states $N_{xz\pm yz}(\omega)$ of identical width. 
Ca induced distortions will enhance these differences, so that the Mott 
transition in these subbands becomes non-trivial. 
This will be also addressed in future studies.               
      
To analyze the Mott transition in Ca$_{2-x}$Sr$_x$RuO$_4$ we have focused
on the variation of the subband occupancies with Coulomb energy. In reality, 
$U$ should be roughly constant as a function of $x$, with $U\approx3.1$~eV 
and $J\approx0.7$~eV according to constrained LDA calculations for $x=2$ 
\cite{pchelkina}. Thus, in Fig.~2 a vertical line near $U=3$~eV should 
qualitatively cover the low temperature phase diagram. $\Delta=0$ with 
$n_{xy}=0.70$, $n_{xz,yz}=0.65$ corresponds to metallic Sr$_2$RuO$_4$. 
$\Delta=0.2$~eV with $n_{xy}=0.76$, $n_{xz,yz}=0.62$ represents the case 
Ca$_{1.5}$Sr$_{0.5}$RuO$_4$, which is also metallic. Finally, the results 
for $\Delta\ge 0.4$~eV indicate that the metal becomes unstable since 
$n_{xy}$ rapidly approaches unity and $n_{xz,yz}$ one-half: the favorable 
electronic configuration for a Mott transition.  

In summary, the paramagnetic metal insulator transition in the layer 
perovskite Ca$_{2-x}$Sr$_x$RuO$_4$ has been investigated within multi-band 
finite temperature ED/DMFT. The results suggest a new concept following 
from the enhanced $d_{xy}$ occupancy induced by Ca doping. Instead of 
orbital-selective Mott transitions, we find a common transition where the 
$d_{xy}$ is completely filled and the remaining half-filled $d_{xz,yz}$ 
bands undergo a standard metal insulator transition. In the pure Sr compound 
strong orbital fluctuations preclude this transition, despite the narrow width 
of the $t_{2g}$ subbands. Thus, realistic Coulomb energies give rise only to a 
weak $d_{xz,yz} \rightarrow d_{xy}$ charge transfer. This transfer is
enhanced by the structural changes due to Sr$\rightarrow$Ca substitution.
Accordingly, orbital fluctuations are reduced and the material becomes 
highly sensitive to local correlations. In the Ca rich compound orbital 
fluctuations are sufficiently weak that the Mott transition occurs 
at realistic values of $U$. On the basis of this picture it would be very 
interesting to study the region $x\le0.5$ more closely in order to understand 
the orbital selective mass enhancement \cite{lee} and the rich magnetic 
phases of this material \cite{nakatsuji}.    
 
One of us (A. L.) likes to thank Theo Costi, Eric Koch and Eva Pavarini
for useful discussions, and E.W. Plummer for comments.\\
Email: a.liebsch@fz-juelich.de; ishida@chs.nihon-u.ac.jp


\begin{thebibliography}{99}

\bibitem{nakatsuji}   
   S. Nakatsuji and Y. Maeno,
       Phys. Rev. Lett. {\bf 84}, 2666 (2000).
   S. Nakatsuji {\it et al.}, 
       {\it ibid.} {\bf 90}, 137202 (2003);
       {\it ibid.} {\bf 93}, 146401 (2004);
   M. Braden {\it et al.}, 
       Phys. Rev. B {\bf 58}, 847 (1998);
   O. Friedt {\it et al.}, 
        {\it ibid.} {\bf 63}, 174432 (2001).

\bibitem{wang}   
   S.-C. Wang, H.-B. Yang, A.K.P. Sekharan, S. Souma, H. Matsui, T. Sato,
   T. Takahashi, C. Lu, J. Zhang, R. Jin, D. Mandrus, E.W. Plummer, 
   Z. Wang, and H. Ding, 
   Phys. Rev. Lett. {\bf 93}, 177007 (2004).

\bibitem{lee}   
   J.S. Lee, S.J. Moon, T.W. Noh, S. Nakatsuji, and Y. Maeno,
   Phys. Rev. Lett. {\bf 96}, 057401 (2006).


\bibitem{maeno}   
   Y. Maeno, T.M. Rice, and M. Sigrist,
   Phys. Today {\bf 54}, 42 (January 2001).
   A.P. Mackenzie and Y. Maeno,
   Rev. Mod. Phys. {\bf 75}, 657 (2003).


\bibitem{oguchi}   
   T. Oguchi,
   Phys. Rev. B {\bf 51}, 1385 (1995);
   I.I. Mazin and D.J. Singh,
   Phys. Rev. Lett. {\bf 79}, 733 (1997).

\bibitem{dmft}
   For a review, see:
   A. Georges, G. Kotliar, W. Krauth and M.J. Rozenberg, 
   Rev. Mod. Phys. {\bf 68}, 13 (1996).

\bibitem{liebsch}
    A. Liebsch and A. Lichtenstein, 
       Phys. Rev. Lett. {\bf 84}, 1591 (2000).

\bibitem{anisimov}  
   V.I. Anisimov, I.A. Nekrasov, D.E. Kondakov, T.M. Rice, and M. Sigrist,
       Euro. Phys. J. B {\bf 25}, 191 (2002).

\bibitem{EPL}
   A. Liebsch, EuroPhys. Lett. {\bf 63}, 97 (2003).

\bibitem{sakai} 
   S. Sakai, R. Arita, K. Held, and H. Aoki,
      Phys. Rev. B {\bf 74}, 155102 (2006).

\bibitem{pchelkina} 
   Z.V. Pchelkina, I.A. Nekrasov, Th. Pruschke, A. Sekiyama, S. Suga,
   V.I. Anisimov, and D. Vollhardt,
      cond-mat/0601507.

\bibitem{dai}
   X. Dai, G. Kotliar, and Z. Fang,
      cond-mat/0611075.

\bibitem{nca}
   Th. Pruschke and N. Grewe, 
       Z. Phys. B {\bf 74}, 439 (1989).

\bibitem{fang}   
   Z. Fang, N. Nagaosa, and K. Terakura,  
   Phys. Rev. B {\bf 69}, 045116 (2004);
   Z. Fang and K. Terakura,  
   Phys. Rev. B {\bf 64}, 020509(R) (2001).

\bibitem{ed}
   C.A. Perroni, H. Ishida, and A. Liebsch,
   Phys. Rev. B {\bf 75}, xxx (2007) [cond-mat/0609702]; 
    see also: 
    A. Liebsch, Phys. Rev. Lett. {\bf 95}, 116402 (2005); 
    A. Liebsch and T.A. Costi, Eur. Phys. J. B {\bf 51}, 523 (2006). 

\bibitem{prb70}
   A. Liebsch,
   Phys. Rev. B {\bf 70}, 165103 (2004).

\bibitem{manini}
   N. Manini, G. Santoro, A. Dal Costa and E. Tosatti,
   Phys. Rev. B {\bf 66}, 115107 (2002).

\bibitem{pavarini}
   E. Pavarini, S. Biermann, A. Poteryaev, A.I. Lichtenstein, A. Georges,
   and O.K. Andersen,
   Phys. Rev. Lett. {\bf 92}, 176403 (2005).


\end{thebibliography}
\end{document}